# Electrical detection of spin-polarized current in topological insulator $Bi_{1.5}Sb_{0.5}Te_{1.7}Se_{1.3}$

**Authors:** Tae-Ha Hwang[1], Hong-Seok Kim[1], Hoil Kim[2,3], Jun Sung Kim[2,3], and Yong-Joo Doh[1,†]

**Affiliations:**

[1] Department of Physics and Photon Science, Gwangju Institute of Science and Technology, Gwangju 61005, Korea

[2] Center for Artificial Low Dimensional Electronic Systems, Institute for Basic Science, Pohang 790-784, Republic of Korea

[3] Department of Physics, Pohang University of Science and Technology, Pohang 790-784, Republic of Korea

† Correspondence to: yjdoh@gist.ac.kr

Abstract

Spin-momentum locked (SML) topological surface state (TSS) provides exotic properties for spintronics applications. The spin-polarized current, which emerges owing to the SML, can be directly detected by performing spin potentiometric measurement. We observed spin-polarized current using a bulk insulating topological insulator (TI), $Bi_{1.5}Sb_{0.5}Te_{1.7}Se_{1.3}$, and Co as the ferromagnetic spin probe. The spin voltage was probed with varying the bias current, temperature, and gate voltage. Moreover, we observed non-local spin-polarized current, which is regarded as a distinguishing property of TIs. The spin-polarization ratio of the non-local current was larger than that of the local current. These findings could reveal a more accurate approach to determine spin-polarization ratio at the TSS.

**Main Text:**

**Introduction**

Topological insulators (TIs) are promising quantum materials, which exhibit strong spin-orbit (SO) coupling. The SO coupling preserves the time-reversal symmetry and combines with the broken inversion symmetry at the surface, generating a spin helical topological surface state (TSS).[1-3] The spins of the carriers transported through the TSS are "locked" to their momentum; this phenomenon is called spin-momentum locking (SML).[4, 5] The SML has been experimentally investigated by optical methods.[2, 3, 6-8] Owing to the inherent spin-polarization of the current caused by the SML, TIs attract significant attention as new functional materials for all-electrical spintronic devices, and provide a new degree of freedom to electronic devices.[9, 10] For instance, current induced SO torque driven magnetoresistive random access memory (MRAM) was proposed.[11] But prior to that, it is very desirable to achieve electrical detection of spin-polarized current. C. H. Li[12] and other groups[13, 14] have reported the detection of spin-polarized current in $Bi_2Se_3$ sample.[3] However, $Bi_2Se_3$ tends to be degenerately n-type doped by Se vacancies, which increases the bulk carrier contribution.[15, 16] Therefore, similar experiments were performed using TI materials that have low doping density such as $(Bi_{1-x}Sb_x)_2Te_3$,[16-18] $Bi_2Te_2Se$,[19] and $Bi_{2-x}Sb_xTe_{3-y}Se_y$ (BSTS).[15, 20]

Meanwhile, band bending at the interface between TI and ferromagnetic (tunnel) junction and at the TI surface must be also considered. The materials with strong SO interaction, like TIs, interfacial electric field due to the band bending induces Rashba type SO coupling and Rashba spin splitting arise.[21-23] The Rashba spin splitting provides current-induced spin-polarization, but in the opposite direction with respect to the SML direction of the TSS.[12, 23] Recently, two oppositely spin-polarized currents were electrically detected using a permalloy-$Al_2O_3$ tunneling contacts to strongly n-type doped $BiSbTeSe_2$ flakes and low-carrier-density $BiSbTeSe_2$ flakes, which correspond to TSS- and Rashba-type-dominated spin-polarized currents, respectively.[20] In the meantime, the overall spin-polarization ratio would decreases, even though one of them dominates.

In this study, we intentionally chose ferromagnetic Co contact on $Bi_{1.5}Sb_{0.5}Te_{1.7}Se_{1.3}$, to minimize the effect of interfacial band bending during the spin-dependent potentiometric measurement. The work function $\Phi$ of Co is 5.0 eV[24], while the work function of BSTS is in the range of 4.95 eV to 5.20 eV[25], and it has energy bandgap of ~ 0.3 eV.[26] We used 1-nm-thick native oxide layer[27, 28] to form a tunneling-type contact between BSTS and Co, which is necessary to avoid the conductance mismatch problem.[29] Furthermore, electrical detection of non-local spin-polarized current was achieved, for the first time in our best knowledge, using the same contact method of local measurement. The non-local spin signal is a direct consequence of surface dominant current flow and negligible conduction in its bulk,[30] hence it may provide a more precise method to determine spin-polarization ratio at TSS.

**Methods**

Single crystals of BSTS were grown using the self-flux method using stoichiometric chunks of high-purity starting materials (Bi, Sb, Te, and Se).[31] The mixture was sealed in an evacuated quartz ampoule and heated up to 850 °C, followed by annealing for two days. The mixture was then slowly cooled to 600 °C for a week and kept at 600 °C for additional one week before furnace cooling. The crystallinity and stoichiometry were confirmed by X-ray diffraction and energy dispersive spectroscopy. The BSTS flakes were mechanically exfoliated onto 300-nm-thick thermally oxidized silicon wafer. The thicknesses of the exfoliated BSTS flakes were 55 nm (D1) and 138 nm (D2), respectively. Ferromagnetic (FM) electrodes (30-nm/20-nm-thick Co/Au) were fabricated on top of each flake using electron-beam lithography and electron-beam deposition. Then, nonmagnetic electrodes (150-nm/50-nm-thick Ti/Au) were fabricated using similar approach. Prior to the normal metal deposition, the samples were dipped for 7 s in a buffered oxide etch solution to eliminate the native oxide. An n-doped Si wafer was employed as the bottom gate electrode. The device parameters are summarized in Table 1. Measurements were performed at a base temperature of 3.0 K, in a cryostat, equipped with a 1 T magnet. The system was connected to Keithley 2400 or Yokogawa GS200 for the DC measurement, SR830 lock-in amplifier for the AC measurement, and Keithley 2182A nanovoltmeter.

**Results and Discussion**

To start, the theoretical principles of spin-potentiometric measurements are discussed, by considering both the energy dispersion relation of TIs and the device structure. For the devices at a temperature ($T$) of 3.0 K, the bulk carrier contribution to the current was highly suppressed, and n-type transport behavior was observed in gate response (Supplementary Fig. S1). Therefore, only electrons flowing through the TSS above the Dirac-point were considered. The schematic diagram of TSS with SML property is shown in Fig. 1 (a). For example, an electron that moves along the *x*-direction ($k_x > 0$, $k_y = 0$) is spin-polarized along the –*y*-direction (defined as down-spin, represented as '↓'), and vice versa, owing to the SML. Figure 1 (b) shows both a false-colored scanning electron microscopy (SEM) image of D1 and the measurement configuration. We defined the magnetic field ($B$) along the up-direction as positive ($B > 0$). Next, a simplified model of spin-potentiometric measurement is provided in Fig. 1 (c). As denoted in the model, when a positive current bias is applied between the two outermost electrodes, up-spin electrons would be more populated than down-spin electrons. In other words, the chemical potential of an up-spin electron $\mu_\uparrow$ is higher than that of a down-spin electron $\mu_\downarrow$. If the FM electrode is magnetized along the up-direction ($M_\uparrow$) or down-direction ($M_\downarrow$) by the external magnetic field, the chemical potential of the FM electrode aligns with $\mu_\downarrow$ or $\mu_\uparrow$. Therefore, the measured voltage $V$ between the FM electrode and normal electrode (its right neighbor) is expected to be $V_{M\uparrow}$ or $V_{M\downarrow}$, for $M_\uparrow$ and $M_\downarrow$, respectively.[22]

Figure 1 (d) shows the spin-dependent voltage as a function of the sweeping magnetic field for a local 4-point geometry of the specimen D1 under a constant DC current bias of 10 µA, and a temperature of 3.7 K. As the magnetic field varies in the positive (red curve) and negative (black curve) directions, clear voltage steps were observed at 180 Oe and –180 Oe, respectively, which correspond to the coercive field of the FM electrode (Supplementary Fig.

S2). The direction of the spin signal is consistent with the theoretical model[22] and previous results.[13-20] By reversing the current direction, opposite electron momentum was generated, hence the spin signal was inverted (Fig. 1 (e)). The voltage hysteresis height ($\Delta V = V_{M\uparrow} - V_{M\downarrow}$) is a linear function of the bias current (Fig. 1 (f)). According to the non-equilibrium Green's function (NEGF)-based model for spin potentiometric measurements in TIs, the voltage difference is directly related with the spin-polarization ratio:[23]

$$\Delta V = (V_{M\uparrow} - V_{M\downarrow}) = I_{FM} R_B P_{FM} (\boldsymbol{p} \cdot \boldsymbol{M_u}) \qquad (1)$$

where $I_{FM}$ is the current bias passing beneath the FM electrode, $1/R_B$ is the ballistic conductance of the channel expressed as $1/R_B = \frac{q^2}{h} \frac{k_F W}{\pi}$, $P_{FM}$ is the effective spin-polarization of the FM electrode, $\boldsymbol{p}$ is the degree of spin-polarization along the *y*-axis per unit current, $\boldsymbol{M_u}$ is the unit vector along the FM magnetization direction, $q$ is the elementary charge, $h$ is the Planck constant, $k_F$ is the Fermi wave number, and $W$ is the width of the channel. Using the valves of $k_F \approx 0.1 \text{ Å}^{-1}$ for BSTS,[32] $P_{FM} = 0.4$ for Co[33, 34] and parameters of each device measurement configuration, we obtain $p = 0.036$ for D1 with local geometry. The obtained value of $\boldsymbol{p}$ is significantly smaller than the theoretical spin-polarization ratio of TSS, $p \sim 50 \%$.[35] It is also smaller than the ratios obtained in other experimental reports, ranging from 0.15 to 0.36 for BS,[12-14] 0.5 for BTS,[19] and $0.78 \pm 0.26$ for BST[17] The difference might be caused by the localized spin states at the tunnel barrier,[36-38] accumulation of bulk carriers at the BSTS/Co tunnel contact interface owing to the Fermi-level pinning,[39] non-ideal spin-detection efficiency of the considered tunnel contact[20], material quality of the considered BSTS flakes, etc. Nevertheless, the obtained value is larger than those reported in other studies ($p = 0.005 \sim 0.01$[15, 20]) using BSTS and permalloy electrode, which has work function in the range of 4.80 to 4.83 eV.[25, 40] It is believed that this is caused by the reduced band bending at the interface between BSTS and Co tunnel contact.

In order to analyze the bulk carrier contribution to the spin-voltage, we applied gate bias with different values, which change the chemical potential of BSTS during the magnetoresistance measurement. Figure 2 (a) shows the measured voltage as a function of the magnetic field using a DC current bias of 20 µA, at different gate voltages ($V_g$) of +20 V (black), 0 V (red), –20 V (green), –40 V (blue), and –60 V (cyan); the background was subtracted, the curves were offset to avoid overlap, and the measurement was performed at a temperature of $T$ = 4.5 K. The extracted $\Delta V$ values for each $V_g$ plot are shown in blue in Fig. 2 (b). It can be noticed that $\Delta V$ increases with the decrease of $V_g$; the relationship is approximately consistent with the relationship between the resistance $R$ and $V_g$, as shown in red in Fig. 2 (b). It is well known that the bottom gate bias applied to a BSTS flake that is thicker than tens of nanometers, usually modulates only the bottom surface conductance (not only TSS of the bottom surface, but also the bulk channel).[31, 41] This implies that positive $V_g$ introduces spin-non-polarized bulk carriers and carriers at the bottom surface state with opposite spin helicity than that of the top surface state.[4, 42] This decreases the ratio between the spin-polarization and total current.

We modulated the temperature of the device during spin-potentiometric measurements. Figure 2 (c) shows the measured voltage as a function of the magnetic field, at a DC current bias of 20 µA, for different temperature values of 4.0 K (black), 8.0 K (red), 12 K (green), and 16 K (blue); the background was subtracted, and the curves were offset for clarity. The extracted $\Delta V$ values for each value of the temperature are shown in blue in Fig. 2 (d). The value of $\Delta V$ is constant for $T$ in the range of 4.0 K to 12 K, then it decreases linearly up to the temperature of 16 K (~ 30% decrease). This trend does not correlate well with the small variations in $R$ as a function of $T$ (red curve in Fig. 2 (d)). The origin of this disagreement is not yet fully understood; however, changes in inelastic scattering rate owing to the electron-electron interactions are expected,[31] instead of the contribution of the thermally activated bulk carriers.

In order to further analyze the effect of the electron-electron scattering, the weak-anti-localization (WAL) behavior was examined for the specimen D1. Prior to the discussion of the temperature dependent WAL effect, we present the dependence of the sheet magnetoconductance (MC), $\Delta\sigma_{2D} = \sigma_{2D}(B) - \sigma_{2D}(B = 0)$, as a function of the magnetic field for different angles $(\theta)$, as shown in Fig. 3 (a). The clear WAL effect, which appeared when the magnetic field was perpendicular $(\theta = 90°)$ to the BSTS surface, was suppressed with the change of the angle towards the parallel $(\theta = 0°)$ direction with respect to the surface. If $B\sin\theta$ is employed as the *x*-axis (Fig. 3 (b)), all of the curves almost match with each other, which indicates that the effect depends only on the perpendicular field. Therefore, the WAL effect emerges owing to the two-dimensional (2D) surface of BSTS.[43, 44] The dependence of the sheet MC as a function of the perpendicular magnetic field is shown in Fig. 3 (c) (measurement points are represented by dots), for several values of the temperature in the range of 4.0 K to 35 K. The cusps near zero field were flattened with the increase of the temperature. The quantum correction to the 2D MC can be described using the Hikami–Larkin–Nagaoka (HLN) model:[30, 43, 45]

$$\Delta\sigma_{2D} = \frac{-\alpha e^2}{2\pi^2 \hbar} \left[ \ln\left(\frac{\hbar}{4eL_\varphi^2 B}\right) - \Psi\left(\frac{1}{2} + \frac{\hbar}{4eL_\varphi^2 B}\right) \right] \quad (2)$$

where $\Psi$ is the digamma function, $e$ is the electron charge, $\hbar$ is the Planck's constant divided by $2\pi$, $L_\varphi$ is the phase relaxation length, and $\alpha$ is the WAL coefficient. The fitting (curves in Fig. 3 (c)) of the sheet MC data using the HLN equation reveals the values of $L_\varphi$ and $\alpha$ for each temperature value. The results are re-plotted as a function of the temperature in Fig. 3 (d). The almost constant value of $\alpha$ (~ −1) for different temperatures indicates robust two-channel transports at the top and bottom surfaces.[43] At 4.0 K, $L_\varphi$ ~ 127 nm; it exhibits power-law dependence as a function of the temperature ($L_\varphi \propto T^{-0.51}$). This implies that the inelastic scattering, caused by the electron-electron interactions, increases with the temperature. This in turn increases the dephasing rate[31], which decreases the spin-polarization ratio at the surface. Nevertheless, we are not able to provide a complete explanation of the dependence of $\Delta V$ as a function of $T$ (Fig. 2 (c)). Various factors might have contributions including the dephasing rate and related quantities.[19]

Meanwhile, the carrier transport at the TSS is not localized between the source and drain; the current flows over the entire TI surface (mainly at the top and bottom surfaces, owing to the dimensions of the considered thin BSTS flake).[30] The non-local voltage ($V_{nloc}$) between the FM electrode and left-most non-magnetic electrode changes linearly with the local current bias ($I_{loc}$) between the non-magnetic electrodes, as shown in Fig. 4 (a), which indicates the existence of $I_{nloc}$. By performing numerical simulations on the devices (Supplementary Figs. S3 and S4), we obtained that the non-local current is proportional to the applied local current bias; for $I_{loc} = 20$ μA, it is estimated that $I_{nloc} = $ -682 nA. In order to detect the spin-polarized non-local current, non-local spin-potentiometric measurements were performed. Schematic of D2 for the non-local measurement configuration is shown in the inset of Fig. 4 (a). At a constant positive ($I_{DC} = 20$ μA, Fig. 4 (b)) and negative ($I_{DC} = -20$ μA, Fig. 4 (c)) current bias, by applying an in-plane magnetic field sweep, the non-local voltage was measured. Spin-voltage hysteresis was observed for both polarities of the current, in opposite directions. The measured spin orientation corresponds to the direction of the non-local current and SML feature of the TSS (insets of Figs. 4 (b) and 4 (c)). It can be noticed that the measured $\Delta V_{nloc}$ increases linearly with the increase of the current bias (Fig. 4. (d)). Taking into account that the measured non-local spin voltage includes contributions only from the

TSS, not from the Rashba state, the above observation is a strong evidence of current-induced spin-polarization owing to the TSS.

Using Eq. (1) and setting $I_{FM} = I_{nloc}$, we obtained that the spin-polarization ratio of the non-local current, $\boldsymbol{p}$, is 0.23. Even though the thickness of BSTS of D1 (55 nm) was smaller than that of D2 (138 nm), $\boldsymbol{p}$ of D2 was 6 times larger than that of D1. This occurs as the bulk carriers without SML property and Rashba effect, which generates opposite direction of SML compared to that of the TSS,[21-23] are absent in the non-local geometry of D2. Therefore, D2 exhibits carrier transport only through the TSS.

**Conclusion**

In conclusion, spin-polarized current through the TSS was measured by an electrical method using BSTS as the TI material and Co as the spin probe, to minimize the band bending. The gate-bias-dependent spin-voltage hysteresis $\Delta V$ was attributed to the accumulation of carriers at the bottom surface. The origin of the temperature-dependent voltage hysteresis $\Delta V$ is not yet fully understood; however, it is believed that it is related to the electron-electron scattering rate associated with complex parameters. Moreover, spin-detection in non-local geometry, based on transport through the TSS, not through Rashba or bulk states, was also observed. Therefore, the highly spin-polarized current through the TSS could provide a more accurate approach to evaluating the SML at the TSS and more efficient all-electric spintronic devices.

**Acknowledgement**

This work was supported by the NRF of Korea through the Basic Science Research Program (2018R1A3B1052827), and the SRC Center for Quantum Coherence in Condensed Matter (2016R1A5A1008184). The work at POSTECH was supported by Institute for Basic

**FIGURE CAPTIONS**

**FIG. 1** (a) Schematic illustration of the Dirac-dispersion of the TSS (left), and Fermi-circle that illustrates the SML. (b) False-colored SEM image of D1. The bluish, reddish, and yellowish colors represent the BSTS flake, Co/Au electrode, and Ti/Au layers of the three normal electrodes, respectively. In addition, the measurement setup is illustrated. (c) Electrochemical potential model to describe the voltage hysteresis when $I_{DC} > 0$ for each magnetization direction of the FM electrode. Measured voltage as a function of the sweeping magnetic field at (d) $I_{DC} = 10$ μA and (e) $I_{DC} = -10$ μA. The voltage hysteresis is observed owing to the SML. Insets in (d) and (e) outline the magnetization direction $M$ of the FM electrode, the directions of the bias current $I$, and corresponding spin-polarization $s$, owing to the SML. (f) $\Delta V$ as a function of the current bias.

**FIG. 2** (a) Measured voltage as a function of the sweeping $B$ field for different values of $V_g$, at $I_{DC} = 20$ μA. (b) Left: $\Delta V$ as a function of $V_g$ including error bars. $\Delta V$ increases with the decrease of the gate voltage. Right: $R$ as a function of $V_g$ of D1. (c) Measured voltage as a function of the sweeping $B$ field for different temperatures, at $I_{DC} = 20$ μA. The linear background was subtracted for each curve in (a) and (b), are they were offset by 20 μV. (d) Left: Obtained $\Delta V$ as a function of $T$ including error bars. Right: $R$ as a function of $T$ of D1.

**FIG. 3** Angle-dependent MC as a function of (a) $B$ and (b) $B\sin\theta$ of D1, measured at 4 K. The inset in (a) shows the measurement schematic. (c) Dependence of MC as a function of the perpendicular $B$ field (dots) for various temperatures and corresponding fitting curves (solid

curves) obtained using the HLN equation. (d) Phase coherence length (red) and dimensionality factor $\alpha$ (blue), as a function of $T$, obtained using the HLN fitting in (c). The black curve represents the power-law dependence of the phase coherence length as a function of $T$.

**FIG. 4** (a) Measured $V_{nloc}$ as a function of the local current bias of D2. Inset shows the non-local measurement schematic for D2. Measured $V_{nloc}$ as a function of the sweeping $B$ field in each direction for local current bias of (b) $I_{DC} = 20$ µA and (c) $I_{DC} = -20$ µA. (d) $\Delta V_{nloc}$ as a function of $I_{nloc}$, obtained by performing numerical simulations. The top $x$-axis represents the corresponding applied local current bias.

**Table. 1** Physical parameters of the BSTS devices.

**FIG. 1**

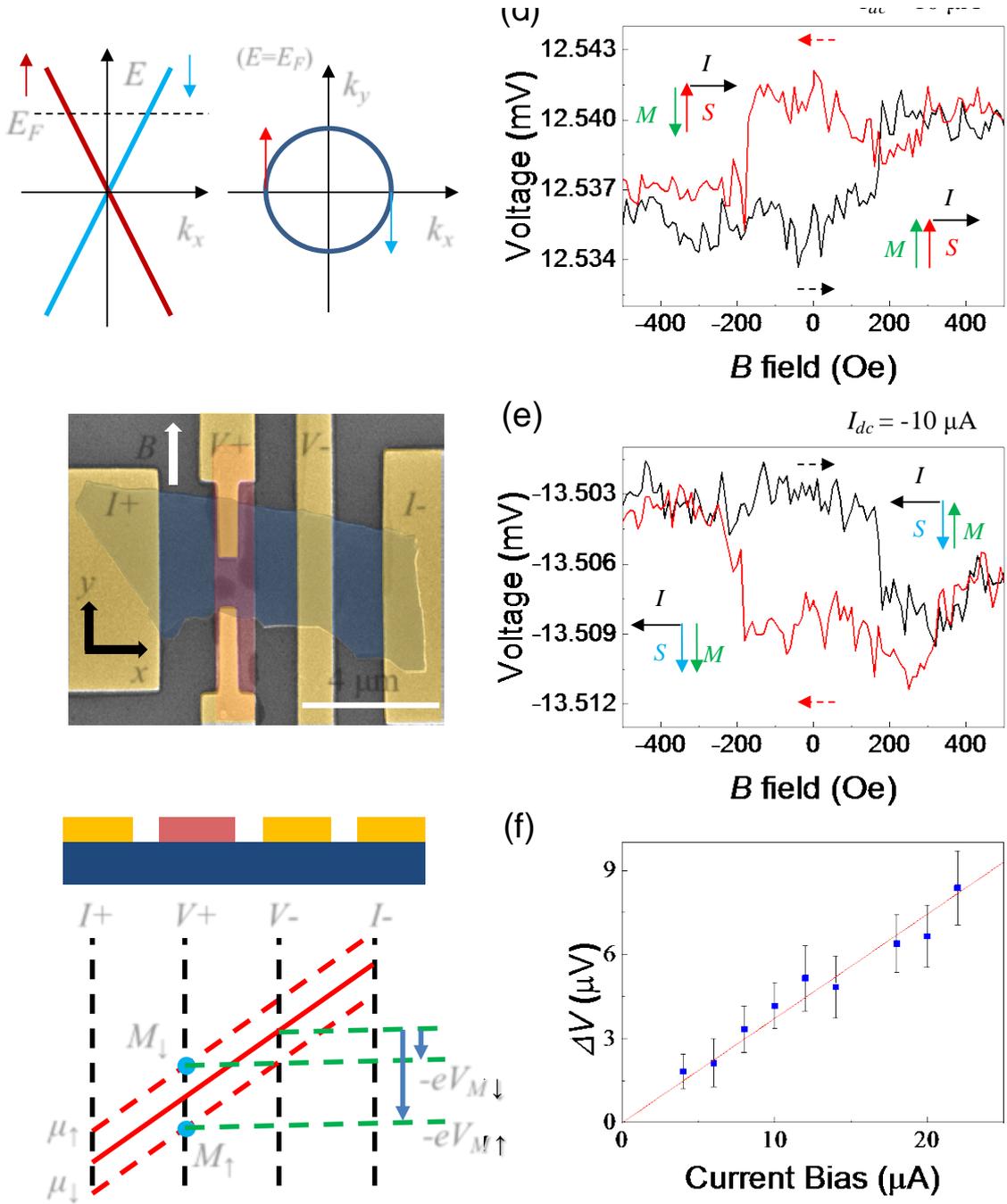

**FIG. 2**

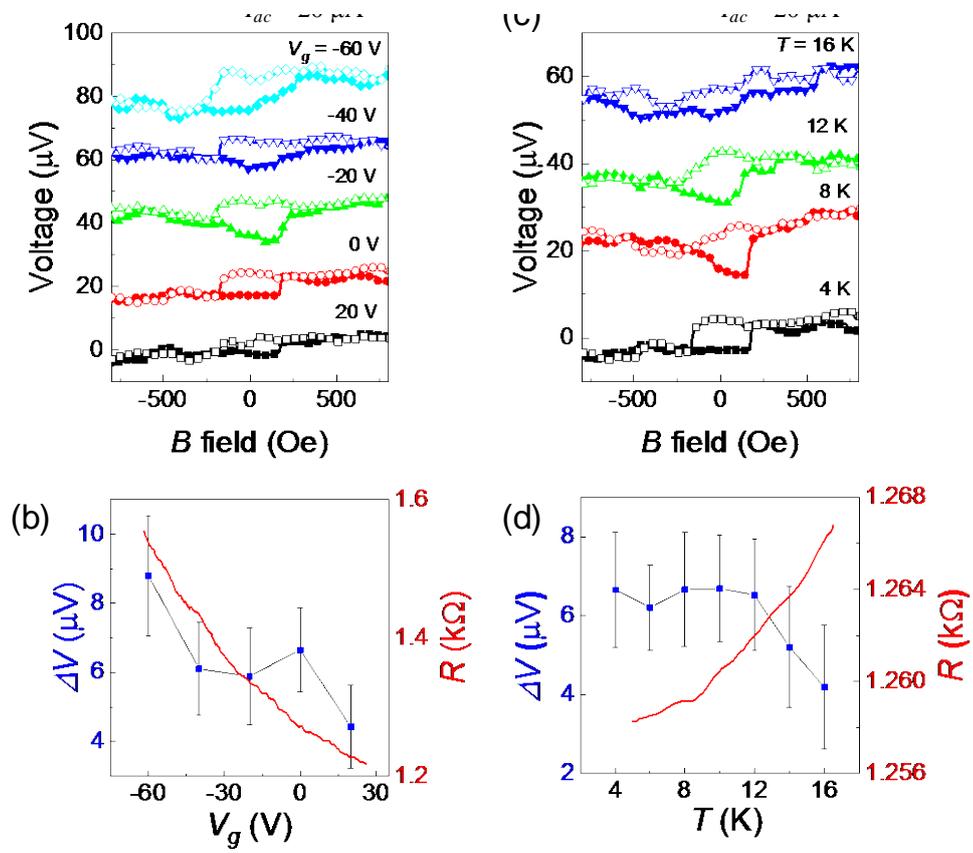

**FIG. 3**

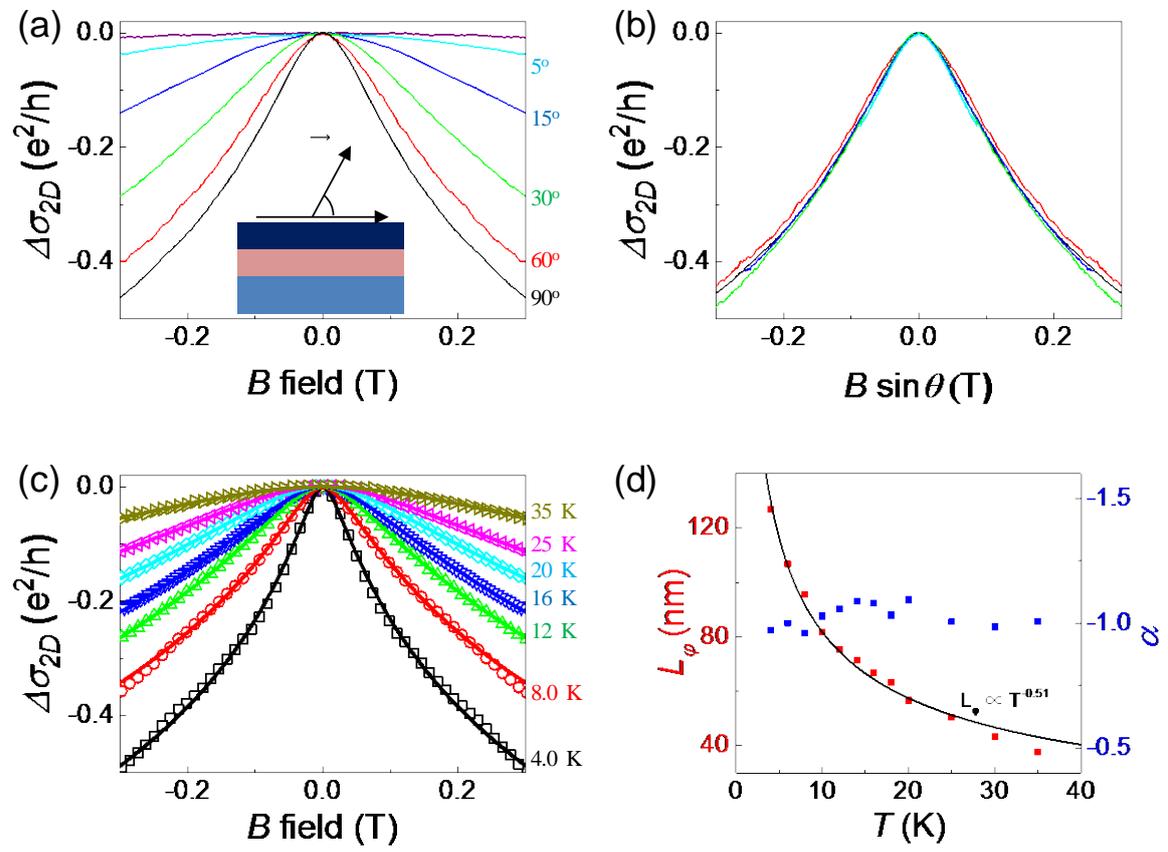

**FIG. 4**

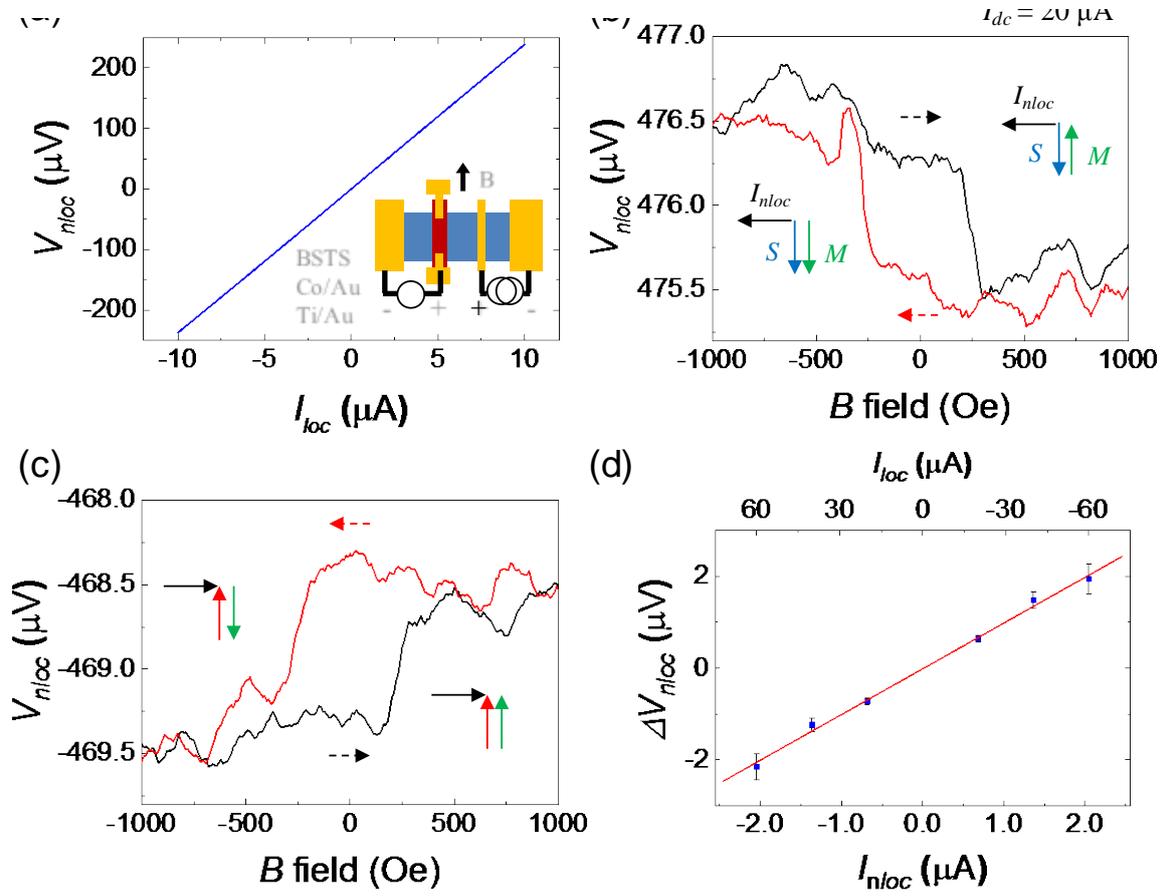

**Table. 1**

| Device No. | Resistivity at 3.7 K (mΩ·cm) | Width of flake (μm) | Thickness of flake (nm) | Electrode spacing (μm) | $H_c$ of Co (Oe) |
|---|---|---|---|---|---|
| **D1** | 19 | 3.4 | 55 | 1.2 | 180 |
| **D2** | 68 | 6.5 | 138 | 1.3 | 220 |
| **D3** | 23 | 4.7 | 86 | 1.3 | - |
| **D4** | 58 | 8.0 | 120 | 1.4 | - |

# Supplementary Information

**Title:** Electrical detection of spin-polarized current in topological insulator $Bi_{1.5}Sb_{0.5}Te_{1.7}Se_{1.3}$

**Authors:** Tae-Ha Hwang[1], Hong-Seok Kim[1], Hoil Kim[2,3], Jun Sung Kim[2,3] and Yong-Joo Doh[1,†]

**Affiliations:**

[1] Department of Physics and Photon Science, Gwangju Institute of Science and Technology, Gwangju 61005, Korea

[2] Center for Artificial Low Dimensional Electronic Systems, Institute for Basic Science, Pohang 790-784, Republic of Korea

[3] Department of Physics, Pohang University of Science and Technology, Pohang 790-784, Republic of Korea

† Correspondence to: yjdoh@gist.ac.kr

1. **Electrical transport characteristics of the BSTS flakes**

Resistivity ($\rho$) vs Temperature ($T$) plot of D1, D2 is shown in Fig. S1 (a). BSTS flake exhibit insulating behavior due to freeze-out of the bulk carriers in high T regime above ~130 K, and metallic behavior in low T regime, indicating topological surface state (TSS) dominates.[1]

Also, gate voltage ($V_g$) dependent $\rho$ at T ~ 4 K is shown in Fig. S1 (b). Generally, $\rho$ keep increase as decreasing $V_g$, signature of n-type transport of BSTS flakes. For D3, $\rho$ increase abruptly near $V_g$ = -75 V and then decrease with decreasing $V_g$, due to ambipolar transport across the dirac point.

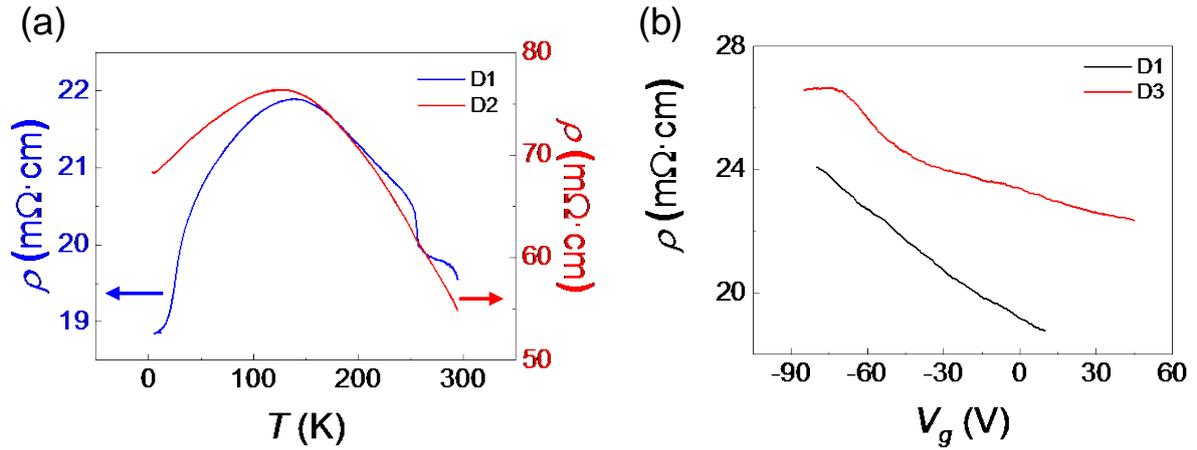

**FIG. S1** (a) Resistivity vs temperature plot for D1 (blue) and D2 (red). (b) Resistivity vs gate voltage plot for D1 (black) and D3 (red) at $T$~ 4 K,

## 2. Longitudinal magnetoresistance of Co/Au electrode

For the ferromagnetic 3d transition metals such as Co, Fe, Ni, most of current is carried by the 4s electrons. Meanwhile, the 3d orbital deforms as the magnetization direction changes, changing s-d scattering probability. As a result, resistance is relatively low when the magnetization direction and the current direction are in perpendicular, while it is high when they are in parallel.[2,3] Therefore, longitudinal magnetoresistance of the Co/Au electrode (D1) in Fig. S2 indicates that magnetization reversal is take place at the resistance dip, in other words, the coercive field ($H_c$) is ±180 Oe.

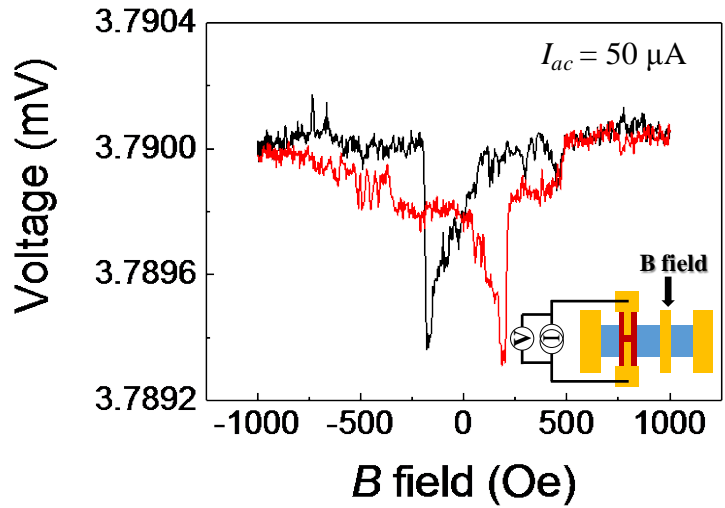

**FIG. S2** Measured voltage vs magnetic (*B*) field of Co/Au electrode of D1 under constant a.c current source of 50 µA at 3.1 K. B field was scanned in positive (red) and negative direction (black).

## 3. Numerical simulations

Current flowing through topological insulator is not localized between source and drain but distributed in entire TSS.[4] To calculate the spin polarization ratio quantitatively, we performed numerical simulation based on finite-element method, using a commercial software of COMSOL.

At first, we carried out measurements and simulations for D4 sample, which has relatively larger surface area of BSTS flake and more number of electrodes than the others, so variety of situations could be tested. Dimension of the BSTS flake were approximated as rectangular plate, and the thickness of TSS is supposed to be 2 nm. Contact resistance was introduced as measured. Simulation model and the measurement configurations for D4 are schematically described in **(Fig. S3 (a))**. We consider four simulation parameters of ($G_{sheet}^{top}$, $G_{sheet}^{bot}$, $\sigma_{side}$, $\sigma_{bulk}$). Here, sheet conductance of Top surface ($G_{sheet}^{top}$) and bottom surface ($G_{sheet}^{bot}$) were set individually, because different amount of band bending is expected at each interface.[4] In addition, each surface structure possibly has its own physical properties, although, gapless state of entire surface of TI is topologically protected. So, side conductivity ($\sigma_{side}$) were set apart from top/bottom surface.[5, 6] Finally, we should consider conductivity of insulating bulk ($\sigma_{bulk}$).

As mentioned, topological insulator has two conducting channels of top and bottom surfaces, so a single current-voltage characteristic cannot represent all the two parameters. If we measure and simulate various geometries with different channel length ($L_{sd}$), we can decide an exact set of ($G_{sheet}^{top}$, $G_{sheet}^{bot}$). At this stage, ($\sigma_{side}$, $\sigma_{bulk}$) were not considered yet. Because side surface cover relatively tiny portion of surface, and BSTS flake is almost bulk-insulating, contribution of ($\sigma_{side}$, $\sigma_{bulk}$) to local current density is negligible. **(Fig. S3 (c))** shows diverse sets of ($G_{sheet}^{top}$, $G_{sheet}^{bot}$) which are in accordance with measurements for different channel lengths between source and drain electrodes ($L_{sd}$), and these three lines converge at a point ($G_{sheet}^{top}$, $G_{sheet}^{bot}$) = (5.9 ± 0.05 $e^2/h$, 2.2 ± 0.1 $e^2/h$).

Once ($G_{sheet}^{top}$, $G_{sheet}^{bot}$) were decided, non-local current distribution depends on ($\sigma_{bulk}$, $\sigma_{side}$). Using parameters of ($\sigma_{bulk}$, $\sigma_{side}$) in **(Fig. S3 (d))**, we get reasonable simulation result of $V_{nloc1}$ and $V_{nolc2}$ at a time, within error range of ± 15 %. Even though their exact ratio may not be estimated, measured and simulated $V_{nloc1}$ and $V_{nolc2}$ values are directly related with non-local current. The current distribution on top and bottom surfaces, and xy cross-sectional view of bulk is visualized in **(Fig. S3 (b))** for ($G_{sheet}^{top}$, $G_{sheet}^{top}$, $\sigma_{side}$, $\sigma_{bulk}$) = (5.91 $e^2/h$, 2.17$e^2/h$, 500 S/m, 0.7 S/m). As expected, most of the current is localized between source and drain electrodes, but some portion of the current spreads out.

Then, the parameters were partially modified and applied to simulating D2 device **(Fig. S4 (a))**, and confirmed that the result is well matched when ($G_{sheet}^{top}$, $G_{sheet}^{top}$, $\sigma_{side}$, $\sigma_{bulk}$) = (5.91 $e^2/h$, 2.17$e^2/h$, 500 S/m, 0.1 S/m). Expected non-local current ($I_{nloc}$) correspond to local current ($I_{loc}$) is graphed in **Fig. S4 (b)**. For a constant current bias of Potential profile and current profile respect to x-position are also shown in **Fig. S4 (c) and (d)**, respectively, for dc current bias of 10 μA. Based on the simulation, we can postulate that $I_{nloc}$ is in negative linear relation with $I_{loc}$., and the amount is 34.1 nA per 1 μA.

The simulation parameters, except $\sigma_{side}$, are comparable with Lee's result,[4] which report on current distribution on BSTS flake. Lower conductivity of side surface than it of top/bottom surface might result from different electronic structure of each surface structure,[6,7] scattering process at the step edge,[8] and et cetra.

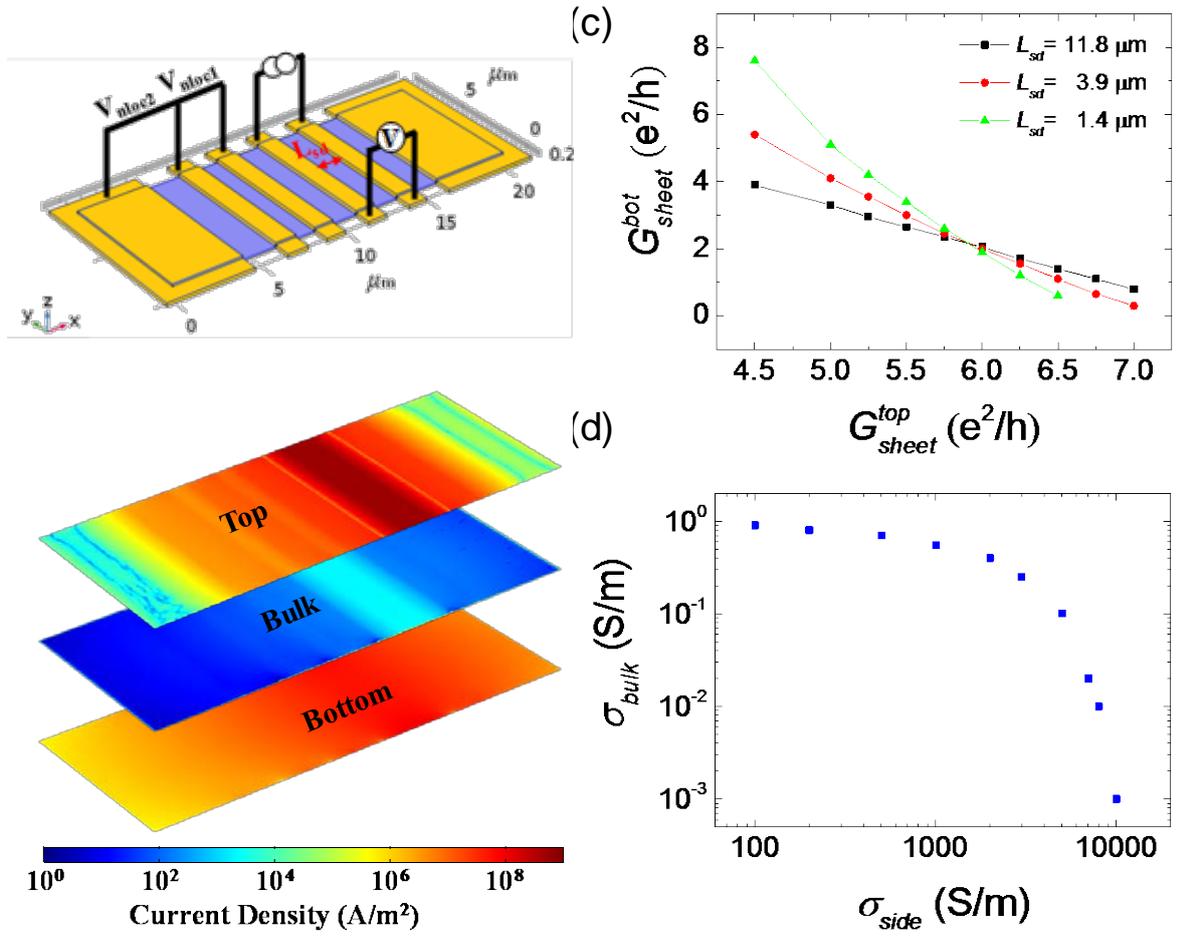

**FIG. S3** (a) schematic view of D4 sample and an example of measurement configuration. (b) color map of current distribution for the top/bottom surface and a bulk cross section. (c) ($G_{sheet}^{top}$, $G_{sheet}^{bot}$) sets to satisfy conductance values for the different channel lengths. (d) ($\sigma_{side}$, $\sigma_{bulk}$) sets to satisfy $V_{nloc1}$ and $V_{nloc2}$.

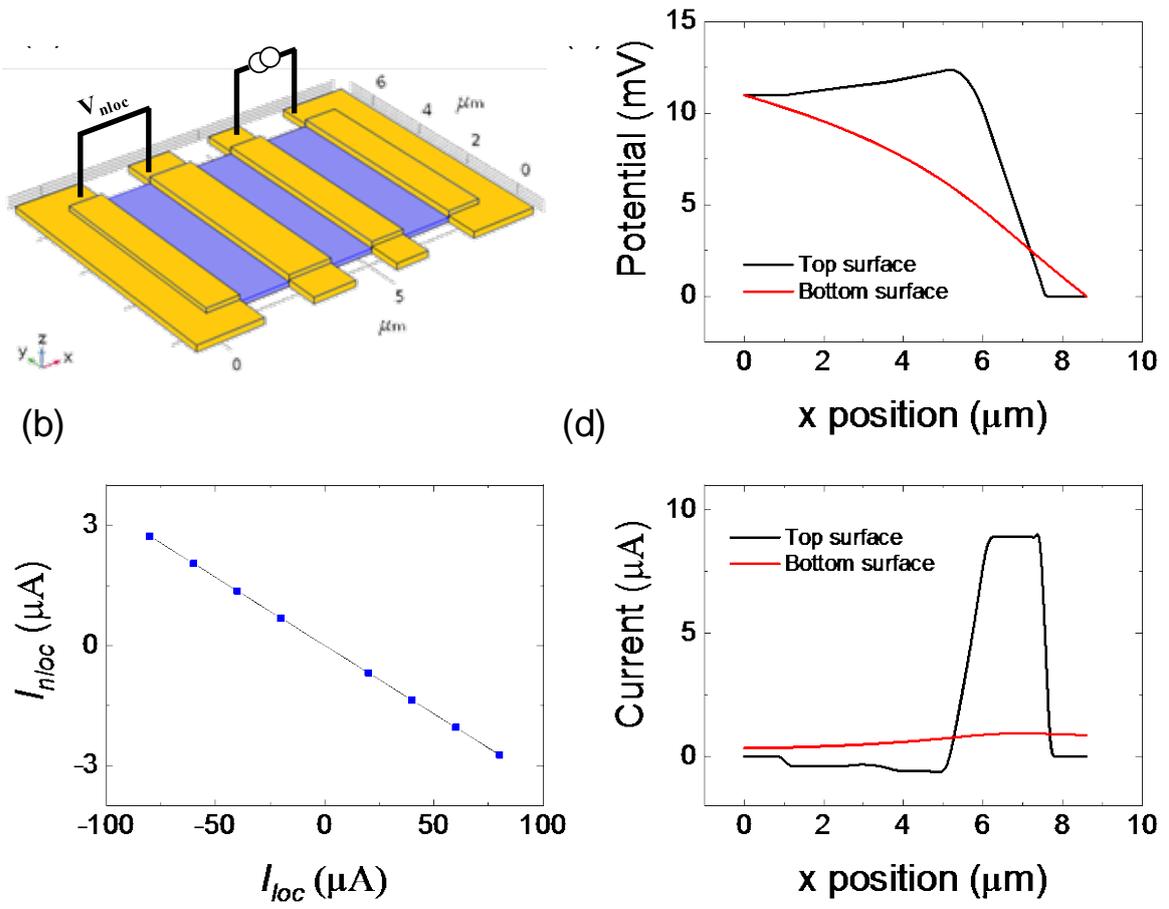

**FIG. S4** (a) schematic view of D2 sample and non-local measurement configuration. (b) expected $I_{nloc}$ vs $I_{loc}$ when ($G_{sheet}^{top}$, $G_{sheet}^{top}$, $\sigma_{side}$, $\sigma_{bulk}$) = (5.91 $e^2/h$, 2.17$e^2/h$, 500 S/m, 0.1 S/m) (c-d) expected potential (c) and current (d) for each of top and bottom surfaces as a function of x position for dc current bias of 10 µA.